\documentclass[aps,pra,twocolumn]{revtex4}
\usepackage{graphicx}
\usepackage{amssymb}
\usepackage{amsmath}
\usepackage{color}

\begin{document}

\title{Revisiting Quantum discord for two-qubit X states: Error bound to Analytical formula}

\author{Min Namkung}
\email{mslab.nk@gmail.com}
\author{Jinho Chang}
\author{Jaehee Shin}
\author{Younghun Kwon}
\email{yyhkwon@hanyang.com}

\affiliation{Department of Physics, Hanyang University, Ansan, Kyunggi-Do, 425-791, South Korea}

\begin{abstract}
 In this article, we investigate the error bound of quantum
discord, obtained by the analytic formula of Ali et al.[Phys. Rev. A
81(2010), 042105] in case of general X states and by the analytic
formula of Fanchini et al.[Phys. Rev. A 81(2010), 052107] in case of
symmetric X states. We show that results of Ali et. al. to general X
states and Fanchini et al. to symmetric X states may have worst-case
error of 0.004565 and 0.0009 respectively.
\end{abstract}


\maketitle


\section{Introduction}
A key ingredient in understanding quantum information may be quantum
correlation. A well-known example of quantum correlation is
entanglement. An entanglement cannot be obtained by a local
operation and classical
communication(LOCC)\cite{m.b.plenio,r.horodecki}. An entanglement is
known to be very fragile to a local noisy channel. Furthermore, it
was shown that a quantum state without entanglement contains
non-locality\cite{c.h.bennett1}. L. Henderson and V.
Vedral\cite{l.henderson} suggested a method to obtain a classical
correlation between parties. H.Ollivier and
W.H.Zurek\cite{h.ollivier} defined a quantum correlation called
quantum discord. The quantum discord can be understood as the
quantum correlation which is the total correlation minus the
classical correlation in a bipartite quantum state. Quantum discord
can contain a value even in a separable state\cite{h.ollivier}. The
quantum discord is invariant under unitary
operation\cite{l.henderson,h.ollivier}. In addition, quantum discord
seems to have a relation with noisy teleportation, entanglement
distillation and quantum state merging\cite{v.madhok}.
Some experimental effort to check quantum discord are under progress\cite{b.p.lanyon, m.gu}.\\
 \indent Quantum discord depends on the measurement
setting. In \cite{h.ollivier} a projective measurement was used in obtaining
quantum discord. The measurement for optimal quantum discord should
build the maximum of the classical correlation. It was shown in \cite{s.hamieh}
that 2 element optimal POVM should be projective measurements.
Quantum discord for a Bell-diagonal state was analytically
obtained\cite{s.luo}. Much research has had a focus on finding
the quantum discord of the X state\cite{m.ali,q.chen,m.shi1,m.shi2,y.huang,x.m.lu,f.f.fanchini}.\\
 \indent There are two reasons why the quantum discord of the two qubit X states becomes significant.
The first reason is that the X state can be obtained through a
unitary transformation to a two qubit state. The second one is that
the general two qubit state can become a X state under a noisy
channel\cite{j.maziero}. Ali et. al. \cite{m.ali} tried to find an
analytic formula of quantum discord to X state. However, it is known
that the result of \cite{m.ali} holds only to special X
state\cite{q.chen}. Fanchini et. al. \cite{f.f.fanchini} tried to
find an analytic formula of the quantum discord for symmetric X
state. Recently, by using von Neumann measurement Y. Huang showed
that the result of \cite{m.ali} may be valid
with some worst-case error\cite{y.huang}. \\
 \indent In this report we show that three element POVM can provide a better quantum discord. Shi et. al\cite{m.shi1,m.shi2} showed that there are some quantum states where 3 element POVM should be used for optimal quantum discord.
The optimality for 3 element POVM can be found by a triangle formed
by the direction vectors. By using 3 element POVM, we numerically
obtained the quantum discord to the quantum state considered in
\cite{y.huang,x.m.lu} and compared it with the result of
\cite{y.huang}.
\section{Quantum Discord}

 The total correlation between the classical subsystem A and B can be
defined by $I(A:B)=H(p^A)+H(p^B)-H(p^A,p^B)$. Here $H(p^X)$($X$ can
be $A$ or $B$) is the Shannon entropy of subsystem $X$. If the
probability distribution of the subsystem becomes
$p^X=\{p_1^X,p_2^X,\cdots,p_n^X\}$, the Shannon entropy is found to
be $H(p^X)=-\sum_{i=1}^n p_i^X \log_2 p_i^X$. $H(p^A,p^B)$ is the
joint entropy of the total system composed of subsystem $A$ and $B$.
When the probability distribution of the total system is known as
$\{p_{ij}^{AB}\}(i=1,2,\cdots,n,j=1,2,\cdots,m)$, joint entropy is
found to be $H(p^A,p^B)=-\sum_{i,j=1}^{n,m}p_{ij}^{AB} \log_2
p_{ij}^{AB}$.

 Let us consider the quantum case. In quantum information one may
consider the information of the quantum system as the quantum state
corresponding to the system. Let $\rho^{AB}$ denote the quantum
state to total system. Then the quantum states of subsystem $A$ and
$B$ can be found by $\rho^A =\mathrm{Tr}_B \rho^{AB}$ and $\rho^B=\mathrm{Tr}_A
\rho^{AB}$. The Von Neumann entropy of $X$ and the total subsystem
are given by $S(\rho^X)=-\mathrm{Tr}\{\rho^X \log_2 \rho^X \}$($X$ becomes
$A$ or $B$) and $S(\rho^{AB})=-\mathrm{Tr}\{\rho^{AB} \log_2 \rho^{AB}\}$
respectively. Therefore the total correlation of the quantum case is
expressed by
\begin{eqnarray} 
I(A:B)=S(\rho^A)+S(\rho^B)-S(\rho^{AB}).
\end{eqnarray}
Total correlation given by Eq(1) contains the classical and quantum
correlation.
  Therefore in order to extract the quantum correlation, one has to subtract the classical correlation from the total correlation.
When one considers the positive operator valued measurement(POVM)
$\{M_{k}^{B}\}$ on subsystem $B$, the classical correlation can be
defined by Eq.(2)\cite{l.henderson}
\begin{eqnarray}
J(A|\{M_k^B\})=S(\rho^A)-\min_{\{M_k^B\}}{\sum_{k} p_k S (\rho_k^A)}.
\end{eqnarray}

Here $\rho_k^A$ is the state of subsystem $A$, given as $\rho_k^A
=\mathrm{Tr}_B \{(1 \otimes M_k^B )\rho^{AB}\}.  S(A|\{M_k^B\})=\sum_k p_k
S(\rho_k^A)$ is the conditional entropy after measurement of
subsystem $B$. Therefore the quantum correlation between subsystem
$A$ and $B$ becomes\cite{h.ollivier}
\begin{eqnarray}
\delta_{\{M_k^B\}}(A:B)=I(A:B)-J(A|\{M_k^B\})\nonumber\\
=S(\rho^B)-S(\rho^{AB})+\min_{\{M_k^B\}}{\sum_k p_k S(\rho_k^A)}.
\end{eqnarray}
This implies that optimizing Eq.(3) is identical to find a
measurement to minimize the conditional entropy $S(A|\{M_k^B\})$.
$S(A|\{M_k^B\})$ is unitary invariant\cite{l.henderson}.

 The X state which appears in various physical cases\cite{j.s.pratt,s.bose} is known
to persist under local noisy channel\cite{t.yu}. The Bell diagonal state
and the Werner state\cite{r.werner} belong to the X state. The general form of
subsystem $\rho^{AB}$ in two qubit states becomes
\begin{eqnarray}
\rho^{AB}&=&\frac{1}{4} \{\mathcal{I}\otimes\mathcal{I}+\sum_i (A_i \mathcal{I}\otimes\sigma_i +B_i \sigma_i\otimes\mathcal{I})\nonumber\\
&&+\sum_{i,j}t_{ij}\sigma_i\otimes\sigma_j\}.
\end{eqnarray}
Here $\sigma_i(i=1,2,3)$ is Pauli's spin matrices and $\vec{A}=(A_1,A_2,A_3), \vec{B}=(B_1,B_2,B_3)$ and $t_{ij}$ can be found by
\begin{eqnarray}
A_i=\mathrm{Tr}\{(\mathcal{I}\otimes\sigma_i)\rho^{AB}\}\nonumber
,\\B_i=\mathrm{Tr}\{(\sigma_i\otimes\mathcal{I})\rho^{AB}\}
,\\t_{ij}=\mathrm{Tr}\{\sigma_i\otimes\sigma_j\}\rho^{AB}\}.\nonumber
\end{eqnarray}
Without loss of generality, one may assume that all the parameters
are real. By applying unitary operations to Eq.(4), one can get the
X state $\rho_X^{AB}$
\begin{eqnarray}
\rho_X^{AB}=\left(\begin{array}{cccc}a&0&0&\epsilon\\
0&b&\delta&0\\0&\delta&c&0\\ \epsilon&0&0&d \end{array} \right).
\end{eqnarray}
  Here, every element of $\rho_X^{AB}$ is real. $a,b,c$ and $d$ satisfies $a+b+c+d=1$.  Eq. (6) can be expressed as
\begin{eqnarray} 
\rho_X^{AB}&=&\frac{1}{4}(\mathcal{I}\otimes\mathcal{I}+A\mathcal{I}\otimes\sigma_3+B\sigma_3\otimes\mathcal{I}\nonumber
\\&&+\sum_i
t_i\sigma_i\otimes\sigma_i).
\end{eqnarray}
Here $A,B,t_1,t_2$ and $t_3$ in Eq. (7) become
\begin{eqnarray} 
A&=&a-b+c-d,\nonumber\\
B&=&a+b-c-d,\nonumber\\
t_1&=&2(\delta+\epsilon),\\
t_2&=&2(\delta-\epsilon),\nonumber\\
t_3&=&a-b-c+d.\nonumber
\end{eqnarray}
\section{Optimization Strategy}
  One may ask whether there exists an optimal POVM for conditional entropy compared to projective measurement.
In \cite{s.hamieh} it was shown that projective measurement is the optimal
condition for the 2 element POVM. Therefore, one should consider
more than two elements POVM. In \cite{m.shi2}, they found that there is X
state where 3 element POVM can be optimal. As is well known, it is
very difficult to handle the optimal 3 element POVM analytically.
The general form of 3 element POVM is expressed as \cite{q.chen}
\begin{eqnarray}
M_k^B = \mu_k (\mathcal{I}+\vec{n}^{(k)}\cdot\vec{\sigma}),~k=1,2,3,~\mu_k>0.
\end{eqnarray}
  Here, $\vec{n}^{(k)}$ is the direction vector to $M_k^B$.
Since, $|\vec{n}^{(k)}|=1$, the positivity of $M_k^B$ holds. When
subsystem B is measured by $\{M_k^B\}$, the post-measurement state
$\rho_k^A$ of subsystem A becomes $\rho_k^A=[1+\{t_1
m_x^{(k)}\sigma_1+t_2 m_y^{(k)}\sigma_2
  +(t_3 m_z^{(k)}+B)\sigma_3\}/(1+Am_z^{(k)})]/2$,
where $m_i^{(k)}(i=x,y,z)$ is a component of the $k$th element in
the direction vector $\vec{m}^{(k)}$. The eigenvalues of $\rho_k^A$
are $\{1\pm E(m_x^{(k)},m_y^{(k)},m_z^{(k)})\}/2$. Here
$E(m_x^{(k)},m_y^{(k)},m_z^{(k)})$ is defined by
\begin{eqnarray} 
&&E(m_x^{(k)},m_y^{(k)},m_z^{(k)})\nonumber
\\&=&\frac{\sqrt{(t_1m_x^{(k)})^2+(t_2m_y^{(k)})^2
+(t_3m_z^{(k)}+B)^2}}{1+Am_z^{(k)}}.
\end{eqnarray}
 The probability to obtain outcome $k$ turns out to be
$p_k=\mu_k(1+m_z^{(k)}A)$. Therefore when 3 element POVM is used for
measurement, the conditional entropy $S(A|\{M_k^B\})$ can be found
as
\begin{eqnarray} 
&&S(A|\{M_k^B\})\nonumber
\\&=&\sum_{k=1}^{3}\mu_k (1+Am_z^{(k)})h(E(m_x^{(k)},m_y^{(k)},m_z^{(k)})).
\end{eqnarray}
 Here $h(x)$
 is a function defined as $h(x)=-\frac{1+x}{2}\log_2\frac{1+x}{2}-\frac{1-x}{2}\log_2
\frac{1-x}{2}$. The complete condition to $\{M_k^B\}$ becomes \cite{q.chen}
\begin{eqnarray}
\mu_1+\mu_2+\mu_3&=&1,
\\\mu_1\vec{n}^{(1)}+\mu_2\vec{n}^{(2)}+\mu_3\vec{n}^{(3)}&=&0.
\end{eqnarray}
  When there are three POVM elements, the direction vector for each element forms a triangle.
The shape of this triangle depends on $\mu_1,\mu_2$ and $\mu_3$.
Fig. 1 shows a triangle made by $\vec{n}^{(1)},\vec{n}^{(2)}$ and
$\vec{n}^{(3)}$ in the XY plane. $\theta_{ij}$ denotes the angle
between the direction vectors $\vec{n}^{(i)}$ and $\vec{n}^{(j)}$.
From Eq.(12)-(13) one can obtain three equations to those angles
\begin{eqnarray}
\mu_1+\mu_2\cos\theta_{12}+\mu_3\cos\theta_{13}=0,\nonumber
\\
\mu_1\cos\theta_{12}+\mu_2+\mu_3\cos\theta_{23}=0,
\\
\mu_1\cos\theta_{13}+\mu_2\cos\theta_{23}+\mu_3=0.\nonumber
\end{eqnarray}
  From Eq.(14) the relations between $\theta_{12},\theta_{23},\theta_{13}$ and $\mu_1,\mu_2,\mu_3$ can be given as
\begin{eqnarray} 
\theta_{12}=\cos^{-1} \frac{\mu_3^2-\mu_1^2-\mu_2^2}{2\mu_1\mu_2},\nonumber
\\\theta_{23}=\cos^{-1} \frac{\mu_1^2-\mu_2^2-\mu_3^2}{2\mu_2\mu_3},
\\\theta_{13}=\cos^{-1} \frac{\mu_2^2-\mu_1^2-\mu_3^2}{2\mu_1\mu_3}.\nonumber
\end{eqnarray}
\begin{figure}[!tt] 
\centerline{\includegraphics[scale=0.5]{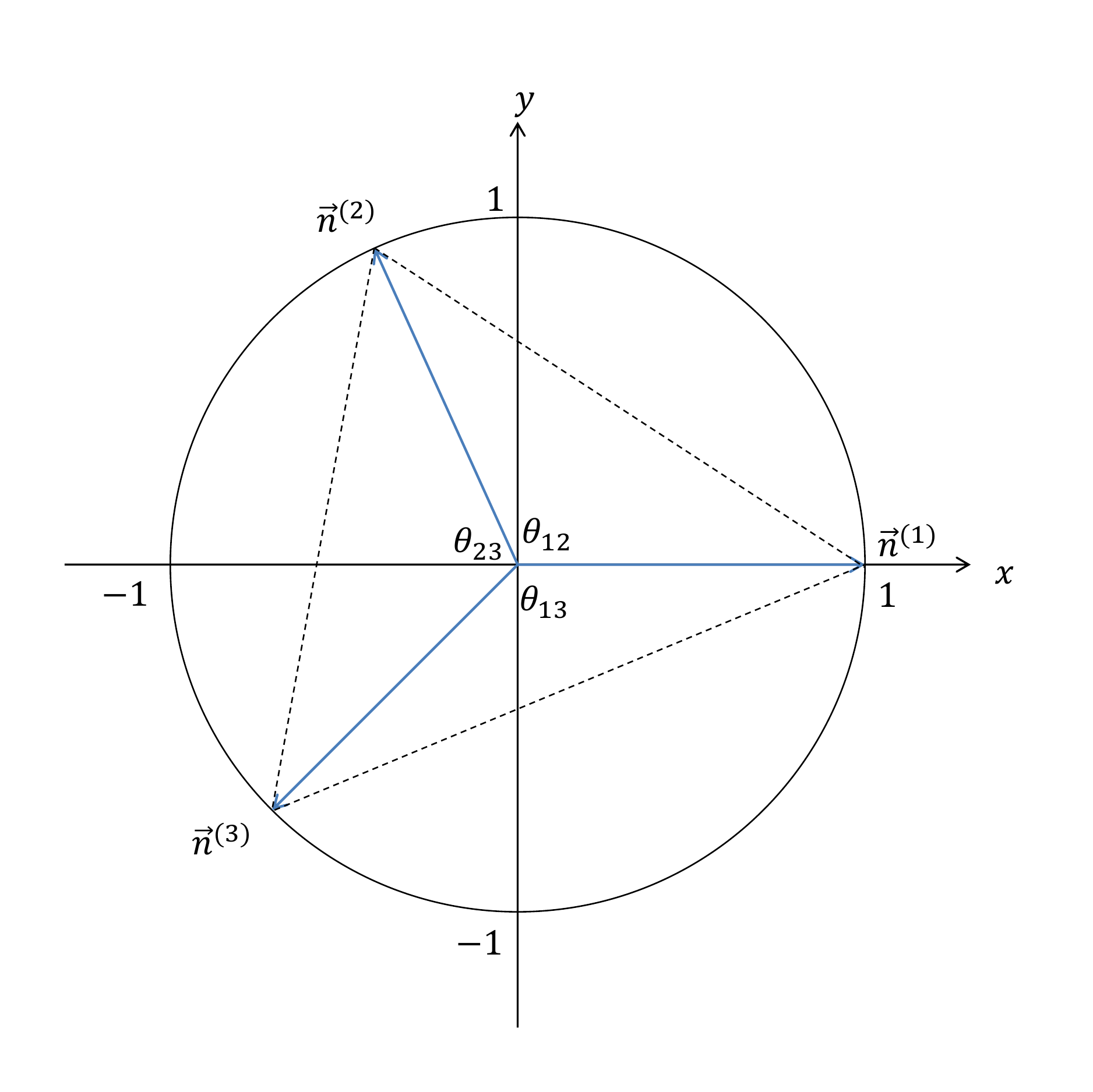}} \caption{A triangle
composed of the direction vector $\vec{n}^{(1)}, \vec{n}^{(2)},
\vec{n}^{(3)}$\label{fig:texgr}}
\end{figure}

  The condition where $\theta_{12},\theta_{23}$, and $\theta_{13}$ are real can be found from $-1<\cos\theta_{12},\cos\theta_{23},\cos\theta_{13}< 1$, which becomes Eq.(16).
Fig. 2 displays the region for $(\mu_1,\mu_2)$ where
$\theta_{12},\theta_{23}$, and $\theta_{13}$ are real.
\begin{eqnarray} 
|\mu_2-\mu_3|<\mu_1<\mu_2+\mu_3\nonumber
\\|\mu_1-\mu_3|<\mu_2<\mu_1+\mu_3
\\|\mu_1-\mu_2|<\mu_3<\mu_1+\mu_2\nonumber
\end{eqnarray}
  However, the measurement illustrated in Fig. 1 does not describe the most general POVM. In order to indicate the most general POVM one has to consider not the XY plane but an arbitrary plane.
\begin{figure}[!tt] 
\centerline{\includegraphics[scale=0.5]{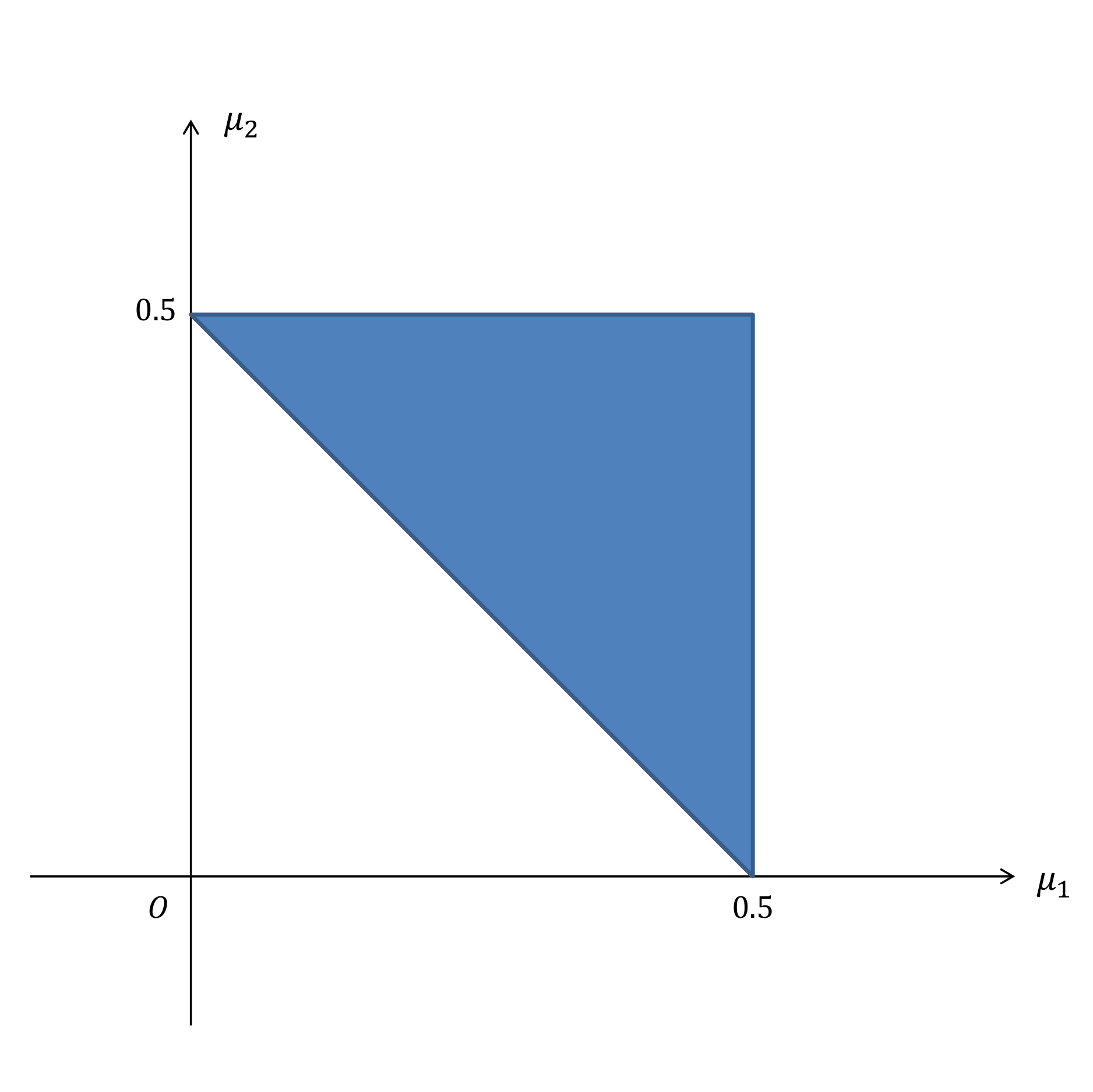}} \caption{The
permitted region of $(\mu_1,\mu_2)$ for 3 element POVM. The edges
are excluded.\label{fig:texgr}}
\end{figure}
  Therefore to find the direction vectors in an arbitrary plane, one can rotate them in Fig. 1 using the Euler angle. The completeness holds under the rotation of the direction vectors. There are three rotation matrices in Eq.(17)
\begin{eqnarray} 
R(\psi,\theta,\phi)=R_{\psi}R_{\theta}R_{\phi},
\end{eqnarray}
  Where each rotation matrices is
\begin{eqnarray} 
R_{\psi}=\left(\begin{array}{ccc}\cos\psi&0&\sin\psi\\
0&1&0\\-\sin\psi&0&\cos\psi \end{array} \right),\nonumber
\\R_{\theta}=\left(\begin{array}{ccc}1&0&0\\
0&\cos\theta&-\sin\theta\\0&\sin\theta&\cos\theta \end{array} \right),
\\R_{\phi}=\left(\begin{array}{ccc}\cos\phi&-\sin\phi&0\\
\sin\phi&\cos\phi&0\\0&0&1 \end{array} \right). \nonumber
\end{eqnarray}
  The direction vector of Fig. 1 are $\vec{n}^{(1)}=(1,0,0), \vec{n}^{(2)}=(\cos\theta_{12},\sin\theta_{12},0), \vec{n}^{(3)}=(\cos\theta_{13},-\sin\theta_{13},0)$. Through rotation matrix $R(\psi,\theta,\phi)$, one can obtain new direction vectors such as
\begin{eqnarray} 
\vec{m}^{(1)}&=&R(\psi,\theta,\phi)\vec{n}^{(1)}\nonumber
\\&=&\left(\begin{array}{c}\cos\phi\cos\psi+\sin\phi\sin\psi\sin\theta\\
\cos\theta\sin\phi\\\sin\phi\cos\psi\sin\theta-\cos\phi\sin\psi\end{array}\right),
\\\nonumber
\vec{m}^{(2)}&=&R(\psi,\theta,\phi)\vec{n}^{(2)}\nonumber
\\&=&\left(\begin{array}{c}\cos(\theta_{12}+\phi)\cos\psi+\sin(\theta_{12}+\phi
)\sin\psi\sin\theta\\
\sin(\theta_{12}+\phi)\cos\theta\\
-\cos(\theta_{12}+\phi)\sin\psi+\sin(\theta_{12}+\phi)\cos\psi
\sin\theta\end{array}\right),
\nonumber
\\\vec{m}^{(3)}&=&R(\psi,\theta,\phi)\vec{n}^{(3)}\nonumber
\\&=&\left(\begin{array}{c}\cos(\theta_{13}-\phi)\cos\psi-\sin(\theta_{13}-\phi
)\sin\psi\sin\theta\\
-\sin(\theta_{13}-\phi)\cos\theta\\
-\cos(\theta_{13}-\phi)\sin\psi-\sin(\theta_{13}-\phi)\cos\psi
\sin\theta\end{array}\right).
\nonumber
\end{eqnarray}
 \indent It is difficult to optimize Eq.(11) analytically. Therefore we use
a Monte-Carlo simulation for optimizing Eq.(11).
 Our strategy is as follows. We randomly select $(\mu_1,\mu_2,\mu_3)$ in the region of Fig. 2. We examine a minimum conditional entropy in the
 region $[0,2\pi]$ to the Euler angle $\psi,\theta,\phi$. It is found that a minimum conditional
 entropy does not depend on $\phi$.

 Y. Huang and Lu et. al considered quantum discord of special X
 states such as
\begin{eqnarray}
\rho_{1}^{AB}=\left(\begin{array}{cccc}0.027180&0&0&0.141651\\
0&0.000224&0&0\\0&0&0.027327&0\\ 0.141651&0&0&0.945269 \end{array}
\right)
\\\rho_{2}^{AB}=\left(\begin{array}{cccc}0.021726&0&0&0.128057\\
0&0.010288&0&0\\0&0&0.010288&0\\ 0.128057&0&0&0.957698 \end{array}
\right)
\end{eqnarray}

and

\begin{eqnarray}
\rho_{3}^{AB}=\left(\begin{array}{cccc}0.0783&0&0&0\\
0&0.1250&0.1000&0\\0&0.1000&0.1250&0\\ 0&0&0&0.6717 \end{array}
\right).
\end{eqnarray}

 In ref.\cite{y.huang} they treated the quantum discord of the X state using projective
measurement. One can see that the condition to the maximum of
discord turns out to be $\theta=\pi/2$. Quantum discords for the X
state of Eq.(20), Eq.(21)\cite{y.huang} and Eq.(22)\cite{x.m.lu} does not change
dramatically according to the measurement setting.

\begin{table}
\centering
\begin{tabular}{c|c|c|c}\hline
 X state&$~~\delta_{3,\min}~~$&$~~\delta_{2,\min}~~$&$~~~\delta_{2}~~~$\\
 \hline
 $\rho_{1}^{AB}$&0.123010&0.124623&0.127575\\
 $\rho_{2}^{AB}$&0.107873&0.107948&0.108773\\
 $\rho_{3}^{AB}$&0.132730&0.132741&0.132751\\
 \hline
\end{tabular} 
\caption{Quantum discord for the states shown in Eq.(20)-(22) when 3
element POVM($\delta_{3,min}$), projective
measurement($\delta_{2,min}$) and measurement obtained by Ali et
al.($\delta_2$) are used respectively. $(\mu_1,\mu_2,\mu_3)$ which
minimizes the quantum discord can be found at (0.4209,0.2938,0.2853)
for $\rho_{1}^{AB}$, (0.4663,0.2489,0.2848) for $\rho_{2}^{AB}$ and
(0.2748,0,2853,0.4349) for $\rho_{3}^{AB}$ respectively.}
\begin{tabular}{c|c|c}\hline
X state&$\Delta_3(\delta_{3,\min}-\delta_2)$&$\Delta_{2}(\delta_{2,\min}-\delta_2)$\\
\hline
$\rho_{1}^{AB}$&$-0.004565$&$-0.002952$\\
$\rho_{2}^{AB}$&$-9.0030\times10^{-4}$&$-8.2542\times10^{-4}$\\
$\rho_{3}^{AB}$&$-2.1109\times10^{-5}$&$-9.6477\times10^{-6}$\\
\hline
\end{tabular}
\caption{Difference  between the quantum discord of 3 element POVM(
2 projective measurement) and that of Ali et al., which
 is denoted by $\Delta_3$($\Delta_2$). }
\end{table}

  Table I shows the quantum discord
of 3 element POVM, that of 2 projective measurements and that of
measurement obtained by Ali et al.($\delta_2$) respectively. As we
can see, $\delta_{3,\min}$ for $\rho_{1}^{AB}$ becomes 0.123010
which is 0.001613 less than the minimum value of the quantum discord
obtained from 2 projective measurement in \cite{y.huang}(The quantum
discord obtained here is a little different from the result of
\cite{y.huang}. It is because the quantum discord obtained in
ref.\cite{y.huang} were expressed in terms of the natural
logarithm($log_{e}$). In this paper Every results to quantum discord
are obtained in terms of $log_{2}$. The results to quantum discord
in terms of the natural logarithm($log_{e}$) can be found in
Appendix.) In addition, $\delta_{3,\min}$ for $\rho_{2}^{AB}$
becomes 0.107873 which is 0.000075 less than the minimum value of
the quantum discord obtained from 2 projective measurement in
\cite{y.huang}. Furthermore $\delta_{3,\min}$ for $\rho_{3}^{AB}$
becomes 0.132730 which is $0.000011$ less than the minimum value of
the quantum discord obtained from 2 projective measurement. For 3
element POVM, the optimized values to $(\mu_1,\mu_2,\mu_3)$ turn out
to be (0.4209,0.2938,0.2853) for $\rho_{1}^{AB}$,
(0.4663,0.2489,0.2848) for $\rho_{2}^{AB}$ and
(0.2748,0,2853,0.4349) for $\rho_{3}^{AB}$ respectively.\\
 Furthermore we can see that a
better bound for $\rho_{1}^{AB}$, $\rho_{2}^{AB}$ and
$\rho_{3}^{AB}$ can be obtained from 3 element POVM. F. Fanchini et.
al provided an analytic formula for the symmetric X state\cite{f.f.fanchini};
however, it is known that the formula may not be optimal. It is
shown that quantum discord to $\rho_{2}^{AB}$ provides a lower value
with 0.000075 than that of F. Fanchini et. al. Table I and II
clearly show that for the quantum states $\rho_{1}^{AB}$,
$\rho_{2}^{AB}$ and $\rho_{3}^{AB}$, 3 element POVM provide a better
value to quantum discord.

\section{Conclusion}
 In this article we investigated the quantum discord to X states considered by
Y. Huang and Lu et. al. We investigated the worst error to quantum
discord from the analytic formula obtained by Ali et al. in case of
general X states and by the analytic formula of Fanchini et al. in
case of symmetric X states. By using  projective measurement Y.
Huang found the worst error to the quantum discord obtained by Ali
et. al. to be 0.002952. In this paper we extend the worst case error
to 0.004565, by using 3 element POVM. Furthermore for symmetric
two-qubit X states, it was found that by using 3 element POVM that
the analytical formula derived by F. F. Fanchini et al. is valid
with worst-case error of 0.0009. In addition, 3 element POVM was
found to supply better quantum discord for the state considered in
Lu et. al. We numerically simulated the lower bound to the quantum
states considered by Y. Huang and Lu et. al. However we still need
to provide an analytic optimal bound for these states, which is in
progress.

\section*{Acknowledgement}
 This work is supported by the Basic Science Research Program through
the National Research Foundation of Korea funded by the Ministry of
Education, Science and Technology (NRF-2010-0025620).

\section*{Appendix. Quantum discord expressed in
terms of the natural logarithm($log_{e}$).}

  In the appendix we supply results to quantum discord in terms of the natural logarithm($log_{e}$).
\begin{table}[b]
\centering
\begin{tabular}{c|c|c|c}\hline
 X state&$~~\delta_{3,\min}~~$&$~~\delta_{2,\min}~~$&$~~~\delta_{2}~~~$\\
 \hline
 $\rho_{1}^{AB}$&0.085264&0.086381&0.088428\\
 $\rho_{2}^{AB}$&0.074772&0.074824&0.075396\\
 $\rho_{3}^{AB}$&0.092001&0.092009&0.092016\\
 \hline
\end{tabular} 
\caption{Revisited quantum discord for the states shown in
Eq.(20)-(22) when 3 element POVM($\delta_{3,min}$), projective
measurement($\delta_{2,min}$) and measurement obtained by Ali et
al.($\delta_2$) are used respectively. The values are expressed in
terms of the natural logarithm($log_{e}$).}
\begin{tabular}{c|c|c}\hline
X state&$\Delta_3(\delta_{3,\min}-\delta_2)$&$\Delta_{2}(\delta_{2,\min}-\delta_2)$\\
\hline
$\rho_{1}^{AB}$&$-0.003164$&$-0.002046$\\
$\rho_{2}^{AB}$&$-6.2400\times10^{-4}$&$-5.7214\times10^{-4}$\\
$\rho_{3}^{AB}$&$-1.4631\times10^{-5}$&$-6.6871\times10^{-6}$\\
\hline
\end{tabular}
\caption{Difference  between the quantum discord of 3 element POVM(
2 projective measurement) and that of Ali et al., which
 is denoted by $\Delta_3$($\Delta_2$). The values are expressed in
terms of the natural logarithm($log_{e}$).}
\end{table}



\begin{thebibliography}{30} 
\bibitem{m.b.plenio}M.~B.~Plenio and S.~Virmani, Quantum Inf. Comput. \textbf{7}(2007), 1
\bibitem{r.horodecki}R.~Horodecki, P.~Horodecki, M.~Horodecki, and K.~Horodecki, Rev. Mod. Phys. \textbf{81}(2009), 865
\bibitem{c.h.bennett1}C.~H.~Bennett, D.~P.~Divincenzo, C.~A.~Fuchs, T.~Mor, E.~Rains, P.~W.~Shor, J.~A.~Smolin, and W.~K.~Wootters,~Phys. Rev. A \textbf{59}(1999), 1070
\bibitem{l.henderson}L.~Henderson and V.~Vedral,~J. Phys. A:Math. Gen. \textbf{34}(2001), 6899
\bibitem{h.ollivier}H.~Ollivier and W.~H.~Zurek,~Phys. Rev. Lett. \textbf{88}(2001), 017901
\bibitem{v.madhok}V.~Madhok, A.~Datta,~Int. J. of Mod. Phys. B, \textbf{27}(2013), 1245041
\bibitem{b.p.lanyon}B.~P.~Lanyon, M.~Barbieri, M.~P.~Almeida and A.~G.~White,~Phys. Rev. Lett. \textbf{101}(2008), 200501
\bibitem{m.gu}M.~Gu, H.~M.~Chrzanowski, S.~M.~Assad, T.~Symul, K.~Modi, T.~C.~Ralph, V.~Vedral, and P.~K.~Lam, Nat. Phys. \textbf{8}(2012), 671
\bibitem{s.hamieh}S.~Hamieh, R.~Kobes, and H.~Zaraket,~Phys. Rev. A \textbf{70}(2004), 052325
\bibitem{s.luo}S.~Luo,~Phys. Rev. A \textbf{77}(2008), 042303
\bibitem{j.maziero}J.~Maziero, T.~Werlang, F.F.~Fanchini, L.C.~Celeri, and R.M.~Serra,~Phys. Rev. A \textbf{81}(2010), 022116
\bibitem{m.ali}M.~Ali, A.R.P.~Rau, and G.~Alber,~Phys. Rev. A \textbf{81}(2010), 042105
\bibitem{q.chen}Q.~Chen, C.~Zhang, S.~Yu, X.X.~Yi, and C.H.~Oh,~Phys. Rev. A  \textbf{84}(2011), 042313
\bibitem{f.f.fanchini}F.F. Fanchini,~T. Werlang,~C. A. Brasil, L. G. E. Arruda, and A. O. Caldeira,~Phys. Rev. A \textbf{81}(2010), 052107
\bibitem{y.huang}Y.~Huang,~Phys. Rev. A \textbf{88}(2013), 014302
\bibitem{m.shi1}M.~Shi, W.~Yang, F.~Jiang, and J.~Du,~J. Phys. A:Math. Theor. \textbf{44}(2011), 415304
\bibitem{m.shi2}M.~Shi, F.~Jiang, X.~Yan, and J.~Du,~Phys. Rev. A \textbf{85}(2012), 064104
\bibitem{x.m.lu}X.M.~Lu, J.~Ma, Z.~Xi and X.~Wang,~Phys. Rev. A \textbf{83}(2011), 012327
\bibitem{j.s.pratt}J. S. Pratt,~Phys. Rev. Lett. \textbf{93}(2004), 237205
\bibitem{s.bose}S. Bose,~I. Fuentes-Guridi,~P. L. Knight, and V. Vedral, Phys. Rev. Lett. \textbf{87}(2001), 050401
\bibitem{t.yu}T. Yu and J. H. Eberly, J. Quant. Inform. and Comput. \textbf{7}(2007), 459
\bibitem{r.werner}R. Werner,~Phys. Rev. A \textbf{40}(1989), 4277





\end{thebibliography}
\end{document}